\documentclass[12pt,letterpaper]{article}
\usepackage{epsfig,rotating,setspace,latexsym,amsmath,epsf,amssymb,bm,theorem}
\usepackage{cite,graphicx,authblk,multirow,color}

\title{Secure Degrees of Freedom of the Multiple Access Wiretap Channel with Multiple Antennas\thanks{This work was supported by NSF Grants CNS 13-14733, CCF 14-22111, CCF 14-22129, and CNS 15-26608, and presented in part at the Asilomar Conference on Signals, Systems and Computers 2015 and to be presented in part at IEEE ICC 2016.}}
\author{Pritam Mukherjee \qquad Sennur Ulukus\\
\normalsize Department of Electrical and Computer Engineering\\
\normalsize University of Maryland, College Park, MD 20742\\
\normalsize {\it pritamm@umd.edu} \qquad {\it ulukus@umd.edu}}

\newtheorem{Theo}{Theorem}

\allowdisplaybreaks

\def\Y{\mathbf{Y}}
\def\X{\mathbf{X}}
\def\Z{\mathbf{Z}}
\def\N{\mathbf{N}}
\def\H{\mathbf{H}}
\def\G{\mathbf{G}}
\def\P{\mathbf{P}}
\def\Q{\mathbf{Q}}
\def\R{\mathbf{R}}
\def\v{\mathbf{v}}
\def\u{\mathbf{u}}
\def\V{\mathbf{V}}

\begin{document}

\maketitle
\begin{abstract}
We consider a two-user multiple-input multiple-output (MIMO) multiple access wiretap channel with $N$ antennas at each transmitter, $N$ antennas at the legitimate receiver, and $K$ antennas at the eavesdropper. We determine the optimal sum secure degrees of freedom (s.d.o.f.) for this model for all values of $N$ and $K$. We subdivide our problem into several regimes based on the values of $N$ and $K$, and provide achievable schemes based on real and vector space alignment techniques for fixed and fading channel gains, respectively.  To prove the optimality of the achievable schemes, we provide matching converses for each regime. Our results show how the number of eavesdropper antennas affects the optimal sum s.d.o.f.~of the multiple access wiretap channel.
\end{abstract}

\section{Introduction}
\begin{figure}
	\centering
	\includegraphics[width=0.6\linewidth]{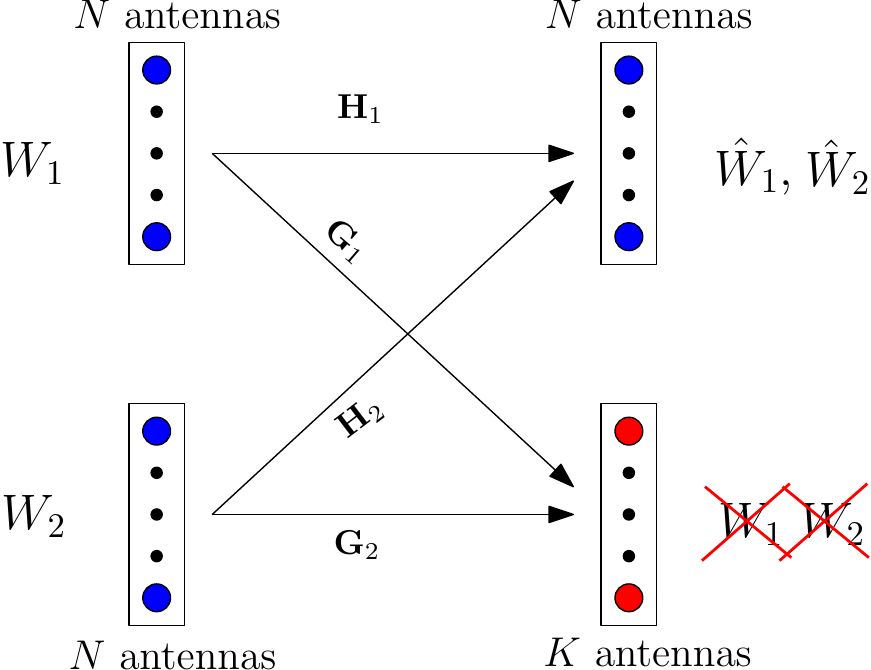}
	\caption{The MIMO multiple access wiretap channel.}
	\label{fig:model}
\end{figure}

We consider the two-user multiple-input multiple-output (MIMO) multiple access wiretap channel where each transmitter has $N$ antennas, the legitimate receiver has $N$ antennas and the eavesdropper has $K$ antennas; see Fig.~\ref{fig:model}. We consider the case when the channel gains are fixed throughout the duration of the communication, as well as the case when the channel is fast fading  and the channel gains vary in an i.i.d.~fashion across time. Our goal in this paper is to characterize how the optimal sum secure degrees of freedom (s.d.o.f.) of the MIMO multiple access wiretap channel varies with the number of antennas at the legitimate users and the eavesdropper. 

To that end, we partition the range of $K$ into various regimes, and propose achievable schemes  for each regime. Our schemes are based on a combination of zero-forcing beamforming and vector space interference alignment techniques. When the number of antennas at the eavesdropper is less than the number of antennas at the transmitters, the nullspace of the eavesdropper channel can be exploited to send secure signals to the legitimate transmitter. This strategy is, in fact, optimal when the number of eavesdropper is sufficiently small ($K\leq \frac{N}{2}$) and the optimal sum s.d.o.f.~is limited by the decoding capability of the legitimate receiver. We note that the optimal scheme requires a single channel use and thus, can be used for both fixed and fading channel gains. 

However, zero-forcing beamforming does not suffice when $K\geq \frac{N}{2}$. In the regime $\frac{N}{2}\leq K \leq \frac{4N}{3}$, the optimal sum s.d.o.f.~is of the form $2\left(d+\frac{l}{3}\right),\, l=0,1,2$, where $d$ is an integer. For the case of fading channel gains, we use vector space interference alignment \cite{cadambe_jafar_interference} over three time slots to achieve the optimal sum s.d.o.f. The structure of the optimal signaling scheme is inspired by ideas from the optimal real alignment scheme presented in \cite{jianwei_ulukus_one_hop} for the single-input single-output (SISO) multiple access wiretap channel. Unlike the previous regime, this scheme for fading channel gains cannot be directly extended to the fixed channel gains case, except for the case  $l=0$, for which the sum s.d.o.f.~is an integer and carefully precoded Gaussian signaling suffices. When $l\neq 0$, the s.d.o.f.~has a fractional part, and Gaussian signaling alone is not optimal. This is also observed in the achievable schemes in \cite{nafea_yener_initial,nafea_yener} for the MIMO wiretap channel with one helper, where structured signaling is used when the optimal s.d.o.f.~is not an integer. However, references \cite{nafea_yener_initial,nafea_yener} consider complex channel gains, for which an s.d.o.f.~of the form $\left(d+\frac{1}{2}\right)$ can be obtained by using $d$ complex symbols (which comprise two real symbols) and one real symbol,  where each real symbol belongs to the same PAM constellation and carries $\frac{1}{2}$ s.d.o.f. In our case, the s.d.o.f.~is of the form $2\left(d+\frac{l}{3}\right),\, l=0,1,2$, and such simplification is not possible even with complex channel gains.

In this paper, we consider real channel gains. In order to handle the fractional s.d.o.f., we decompose the channel input at each transmitter into two parts: a Gaussian signaling part carrying $d$ (the integer part) d.o.f.~of information securely, and a structured signaling part carrying  $\frac{l}{3}$ (the fractional part) d.o.f.~of information securely. The structure of the Gaussian signals carrying the integer s.d.o.f.~resembles that of the schemes for the fading channel gains. When $l=1$, we design the structured signals carrying $\frac{2}{3}$ sum s.d.o.f.~according to the real interference alignment based SISO scheme of \cite{jianwei_ulukus_one_hop}. However, when $l=2$,  a new scheme is required to achieve $\frac{4}{3}$ sum s.d.o.f.~on the MIMO multiple access wiretap channel with two antennas at every terminal. To that end, we provide a novel optimal scheme for the canonical $2\times2\times2\times2$ MIMO multiple access wiretap channel. Interestingly, the scheme relies on asymptotic real interference alignment \cite{real_inter_align_exploit} at  each antenna of the legitimate receiver.

When the number of eavesdropper antennas $K$ is large enough $K\geq \frac{4N}{3}$, the optimal sum s.d.o.f.~is given by $(2N-K)$, which is always an integer. In this regime Gaussian signaling along with vector space alignment techniques suffices. In fact, the scheme uses only one time slot and can be used with both fixed and fading channel gains. When the number of antennas at the eavesdropper is very large ($K\geq \frac{3N}{2}$), the two-user multiple access wiretap channel reduces to a wiretap channel with one helper, and, thus, the scheme for the MIMO wiretap channel with one helper in \cite{nafea_yener} is optimal.

To establish the optimality of our achievable schemes, we present matching converses in each regime. A simple upper bound is obtained by allowing cooperation between the two transmitters. This reduces the two-user multiple access wiretap channel to a MIMO wiretap channel with $2N$ antennas at the transmitter, $N$ antennas at the legitimate receiver and $K$ antennas at the eavesdropper. The optimal s.d.o.f.~of this MIMO wiretap channel is well known to be $\min((2N-K)^+,N)$ \cite{mimo_wiretap,khisti_mimome}, and this serves as an upper bound for the sum s.d.o.f.~of the two-user multiple access wiretap channel. This bound is optimal when the number of eavesdropper antennas $K$ is either quite small ($K\leq \frac{N}{2}$), or quite large ($K\geq \frac{4N}{3}$). When $K$ is small, the sum s.d.o.f.~is limited by the decoding capability of the legitimate receiver, and the optimal sum s.d.o.f.~is $N$ which is optimal even without any secrecy constraints. When $K$ is large, the s.d.o.f.~is limited by the requirement of secrecy from a very strong eavesdropper. For intermediate values of $K$, the distributed nature of the transmitters dominates, and we employ a generalization of the SISO converse techniques of \cite{jianwei_ulukus_one_hop} for the converse proof in the MIMO case, similar to \cite{nafea_yener}.      

\emph{Related Work:} The multiple access wiretap channel is introduced by \cite{tekin-yener-it2,tekin_yener_mac2008}, where the technique of cooperative jamming is introduced to improve the rates achievable with Gaussian signaling. Reference \cite{ersen_ulukus_mac_2008} provides outer bounds and identifies cases where these outer bounds are within 0.5 bits per channel use of the rates achievable by Gaussian signaling. While the exact secrecy capacity remains unknown, the achievable rates in \cite{tekin-yener-it2,tekin_yener_mac2008,ersen_ulukus_mac_2008} all yield zero s.d.o.f. Reference \cite{bassily_ergodic_align} proposes scaling-based and ergodic alignment techniques to achieve a sum s.d.o.f.~of $\frac{K-1}{K}$ for the $K$-user MAC-WT; thus, showing that an alignment based scheme strictly outperforms i.i.d.~Gaussian signaling with or without cooperative jamming at high SNR. Finally, references \cite{jianwei_ulukus_one_hop,jianwei_mac_region_journal} establish the optimal sum s.d.o.f.~to be $\frac{K(K-1)}{K(K-1)+1}$ and the full s.d.o.f.~region, respectively, for the SISO multiple access wiretap channel.  Other related channel models are the wiretap channel with helpers and the interference channel with confidential messages, for which the optimal sum s.d.o.f.~is known for the SISO and MIMO cases in \cite{jianwei_ulukus_one_hop} and \cite{nafea_yener_initial,nafea_yener}, and in \cite{jianwei_interference} and \cite{karim_ulukus_asilomar15}, respectively. 

\section{System Model}
The two-user multiple access wiretap channel, see Fig.~\ref{fig:model}, is described by,
\begin{align}
\Y(t) =& \H_1(t)\X_1(t) + \H_2(t)\X_2(t) + \N_1(t)\\
\Z(t) =& \G_1(t)\X_1(t) + \G_2(t)\X_2(t) + \N_2(t)
\end{align}
where $\X_i(t)$ is an $N$ dimensional column vector denoting the $i$th user's channel input, $\Y(t)$ is an $N$ dimensional vector denoting the legitimate receiver's channel output, and $\Z(t)$ is a $K$ dimensional vector denoting the eavesdropper's channel output, at time $t$. In addition, $\N_1(t)$ and $\N_2(t)$ are $N$ and $K$ dimensional white Gaussian noise vectors, respectively, with $\N_1 \sim \mathcal{N}(\mathbf{0},\mathbf{I}_{N})$ and  $\N_2 \sim \mathcal{N}(\mathbf{0},\mathbf{I}_{K})$, where $\mathbf{I}_N$ denotes the $N\times N$ identity matrix. Here, $\H_i(t)$ and $\G_i(t)$ are the $N\times N$ and $K\times N$ channel matrices from transmitter $i$ to the legitimate receiver and the eavesdropper, respectively, at time $t$. When the channel gains are fixed, the entries of $\H_i(t)$ and $\G_i(t)$ are drawn from an arbitrary but fixed continuous distribution with bounded support in an i.i.d.~fashion prior to the start of the communication, and remain fixed throughout the duration of the communication, i.e., for $1\leq t\leq n$. When the channel gains are fading, the entries of $\H_i(t)$ and $\G_i(t)$ are drawn from the fixed continuous distribution with bounded support in an i.i.d.~fashion at every time slot $t$. We assume that the channel matrices $\H_i(t)$ and $\G_i(t)$ are known with full precision at all terminals, at time $t$. All channel inputs satisfy the average power constraint $E[\lVert \X_i(t)\rVert^2]\leq P,\; i=1,2$, where $\lVert \X \rVert$ denotes the Euclidean (or the spectral norm) of the vector (or matrix) $\X$.

Transmitter $i$ wishes to send a message $W_i$, uniformly distributed in $\mathcal{W}_i$, securely to the legitimate receiver in the presence of the eavesdropper. A secure rate pair $(R_{1},R_{2})$, with $R_{i} = \frac{\log|\mathcal{W}_i|}{n}$ is achievable if there exists a sequence of codes which satisfy the reliability constraints at the legitimate receiver, namely, $\mbox{Pr}[ W_{i}\neq \hat{W}_{i}] \leq \epsilon_{n}$, for $i=1,2$, and the secrecy constraint, namely,
\begin{align}
\frac{1}{n} I(W_{1},W_2;\Z^n) \leq \epsilon_n
\end{align}
where $\epsilon_n \rightarrow 0$ as $n \rightarrow \infty$. An s.d.o.f.~pair  $\left(d_1, d_2 \right) $ is said to be achievable if a rate pair $\left(R_1,R_2 \right) $ is achievable with 
\begin{align}
d_i = \lim\limits_{P\rightarrow \infty} \frac{R_i}{\frac{1}{2}\log P}
\end{align}
The sum s.d.o.f.~$d_s$ is the largest achievable $d_1+d_2$.

\section{Main Result}
The main result of this paper is the determination of the optimal sum s.d.o.f.~of the MIMO multiple access wiretap channel. We have the following theorem.
\begin{Theo}\label{theo}
The optimal sum s.d.o.f.~of the MIMO multiple access wiretap channel with $N$ antennas at the transmitters, $N$ antennas at the legitimate receiver and $K$ antennas at the eavesdropper is given by
\begin{align}
d_s = \begin{cases}
N, &\mbox{if } K\leq \frac{1}{2}N\\
\frac{2}{3}(2N-K), &\mbox{if } \frac{1}{2}N \leq  K\leq N\\ 
\frac{2}{3}N, &\mbox{if } N \leq  K\leq \frac{4}{3}N\\
2N-K, &\mbox{if } \frac{4}{3}N \leq  K\leq 2N\\
0,&\mbox{if }  K\geq 2N.   
\end{cases}\label{eq:sum_dof}
\end{align} 
for almost all channel gains.
\end{Theo}

We present the converse proof for this theorem in Section \ref{sec:converse}. The achievable schemes for the case of fading channel gains are presented in Section \ref{sec:achievability}, while the achievable schemes for the case of fixed channel gains are presented in Section
\ref{sec:achievability_real}. 

\begin{figure}
\centering
\includegraphics[width=0.6\linewidth]{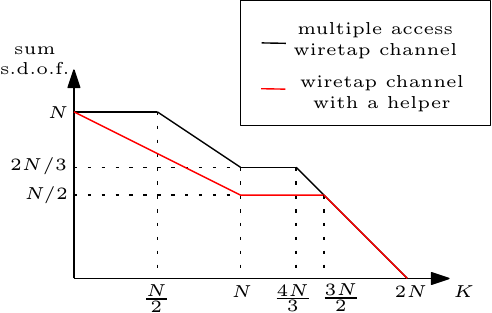}
\caption{$d_s$ versus $K$.}
\label{fig:sdof_plot}
\end{figure}

Fig.~\ref{fig:sdof_plot} shows the variation of the optimal sum s.d.o.f.~with the number of eavesdropper antennas $K$. Note that as in the SISO case, the optimal sum s.d.o.f.~is higher for the multiple access wiretap channel than for the wiretap channel with one helper \cite{nafea_yener}, when $K < 3N/2$. However, when the number of eavesdropper antennas  $K$ is large enough, i.e., when $K \geq 3N/2$, the optimal sum s.d.o.f.~of the multiple access wiretap channel is the same as the optimal s.d.o.f.~of the wiretap channel with a helper.   

Further, note that when the number of eavesdropper antennas $K$ is small enough ($K\leq \frac{N}{2}$), the optimal sum s.d.o.f.~is $N$, which is the optimal d.o.f.~of the multiple access channel without any secrecy constraints. Thus, there is no penalty for imposing the secrecy constraints in this regime. Also note that allowing cooperation beteen the transmitters does not increase the sum s.d.o.f.~in this regime. Heuristically, the eavesdropper is quite weak in this regime, and the optimal sum s.d.o.f.~is limited by the decoding capabilities of the legitimate receiver. 

On the other hand, when the number of antennas $K$ is quite large ($K \geq \frac{4N}{3}$), the optimal sum s.d.o.f.~is $(2N-K)$, which is the optimal s.d.o.f.~obtained by allowing cooperation between the transmitters. Intuitively, the eavesdropper is very strong in this regime and the sum s.d.o.f.~is limited by the requirement of secrecy from this strong eavesdropper. In the intermediate regime, when $\frac{N}{2} \leq K \leq \frac{4N}{3}$, the distributed nature of the transmitters becomes a key factor and the upper bound obtained by allowing cooperation between the transmitters is no longer achievable; see Fig.~\ref{fig:converse_proof}.
    
\section{Proof of the Converse}\label{sec:converse}
We prove the following upper bounds which are combined to give the converse for the full range of $N$ and $K$, 
\begin{align}
d_1+d_2 \leq& \min((2N-K)^+, N) \label{eq:cooperative_bound}\\
d_1+d_2 \leq& \max\left(\frac{2}{3}(2N-K),\frac{2}{3}N\right) \label{eq:jianwei_bound}
\end{align}
where $(x)^+$ denotes $\max(x,0)$.
 
\begin{figure}
\centering
\includegraphics[width=0.7\linewidth]{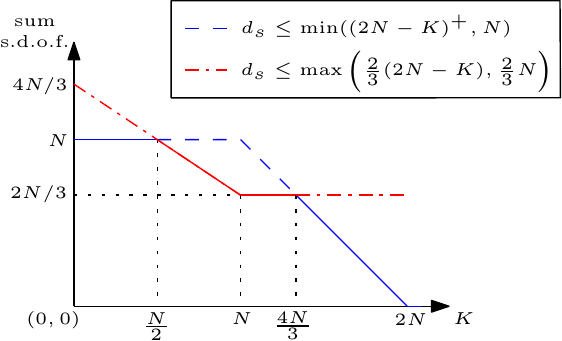}
\caption{The two upper bounds.}
\label{fig:converse_proof}
\end{figure}

It can be verified from Fig.~\ref{fig:converse_proof} that the minimum of the two bounds in \eqref{eq:cooperative_bound}-\eqref{eq:jianwei_bound} gives the converse to the sum s.d.o.f.~stated in \eqref{eq:sum_dof} for all ranges of $N$ and $K$. Thus, we next provide proofs of each of the bounds in \eqref{eq:cooperative_bound} and \eqref{eq:jianwei_bound}.

\subsection{Proof of $d_1+d_2 \leq \min((2N-K)^+, N)$}
This bound follows by allowing cooperation between the transmitters, which reduces the two-user multiple access wiretap channel to a single-user MIMO wiretap channel with $2N$ antennas at the transmitter, $N$ antennas at the legitimate receiver and $K$ antennas at the eavesdropper. The optimal s.d.o.f.~for this MIMO wiretap channel is known to be $\min((2N-K)^+, N)$ \cite{mimo_wiretap,khisti_mimome}.

\subsection{Proof of $d_1+d_2 \leq \max\left(\frac{2}{3}(2N-K),\frac{2}{3}N\right)$}  
We only show that $d_1+d_2 \leq \frac{2}{3}(2N-K)$, when $K\leq N$, and note that the bound $d_1+d_2\leq \frac{2}{3}N$ for $K>N$ follows from the fact that increasing the number of eavesdropper antennas cannot increase the sum s.d.o.f.; thus, the sum s.d.o.f.~when $K>N$ is  upper-bounded by the sum s.d.o.f.~for the case of $K=N$, which is $\frac{2}{3}N$.

To prove $d_1+d_2 \leq \frac{2}{3}(2N-K)$ when $K\leq N$, we follow \cite{jianwei_ulukus_one_hop,nafea_yener}. We define noisy versions of $\X_i$ as $\tilde{\X}_i = \X_i + \tilde{\N}_i$ where $\tilde{\N}_i \sim \mathcal{N}(\mathbf{0},\rho_i^2\mathbf{I}_N)$ with $\rho_i^2 < \min\left(\frac{1}{\lVert \H_i \rVert^2},\frac{1}{\lVert \G_i \rVert^2} \right)$. The \emph{secrecy penalty lemma} \cite{jianwei_ulukus_one_hop} can then be derived as
\begin{align}
n(R_1 + R_2) \leq& I(W_1,W_2;\Y^n| \Z^n) + n\epsilon\\
\leq& h(\Y^n|\Z^n) + nc_1\\
=& h(\Y^n,\Z^n) - h(\Z^n) +nc_1\\
\leq& h(\tilde{\X}_1^n,\tilde{\X}_2^n) - h(\Z^n) + nc_2\\
\leq& h(\tilde{\X}_1^n) + h(\tilde{\X}_2^n) - h(\Z^n) + nc_2 \label{eq:sec_penalty_lemma1}   
\end{align}     
Now consider a stochastically equivalent version of $\Z$ given by $\tilde{\Z} = \G_1\tilde{\X}_1 + \G_2\X_2 + \N_Z$, where $\N_Z$ is an independent Gaussian noise vector, distributed as $\mathcal{N}(\mathbf{0},\mathbf{I}_K - \rho_1^2\G_1\G_1^H)$.      
Further, let $\G_1 = [\tilde{\G}_1\quad \hat{\G}_1]$ and $\tilde{\X}_1^T = [\tilde{\X}_{1a}^T\quad\tilde{\X}_{1b}^T]^T$, where $ \tilde{\G}_1$ is the matrix with the first $K$ columns of $\G_1$, $\hat{\G}_1$ has the last $N-K$ columns of $\G_1$, $\tilde{\X}_{1a}$ is a vector with the top $K$ elements of $\tilde{\X}_1$, while  $\tilde{\X}_{1b}$ has the remaining $N-K$ elements of $\tilde{\X}_1$. Then, we have
\begin{align}
h(\Z^n) =h(\tilde{\Z}^n) =& h(\G_1^n\tilde{\X}_1^n + \G_2^n\X_2^n+\N_Z^n) \\
\geq& h(\G_1^n\tilde{\X}_1^n)\\
=& h(\tilde{\G}_1^n\tilde{\X}_{1a}^n +\hat{\G}_1^n\tilde{\X}_{1b}^n) \\
\geq& h(\tilde{\G}_1^n\tilde{\X}_{1a}^n| \tilde{\X}_{1b}^n) \\
=& h(\tilde{\X}_{1a}^n| \tilde{\X}_{1b}^n) + nc_3 \label{eq:bound_on_z}
\end{align} 
Using \eqref{eq:bound_on_z} in \eqref{eq:sec_penalty_lemma1}, we have
\begin{align}
n(R_1 + R_2) \leq& h( \tilde{\X}_{1b}^n) + h(\tilde{\X}_2^n) + nc_4 \label{eq:sec_penalty_lemma2}
\end{align}

The \emph{role of a helper lemma} \cite{jianwei_ulukus_one_hop} also generalizes to the MIMO case as
\begin{align}
nR_1 \leq& I(\X_1^n;\Y^n)\\
=& h(\Y^n) - h(\H_2^n\X_2^n + \N_1^n)\\
\leq& h(\Y^n) - h(\tilde{\X}_2^n) + nc_5 \label{eq:role_of_helper}
\end{align}

Adding \eqref{eq:sec_penalty_lemma2} and \eqref{eq:role_of_helper}, we have
\begin{align}
n(2R_1+R_2) \leq& h(\Y^n) + h( \tilde{\X}_{1b}^n) + nc_6\\
\leq& N \frac{n}{2}\log P + (N-K)\frac{n}{2}\log P + nc_7\\
=& (2N-K) \frac{n}{2}\log P + nc_7
\end{align}
First dividing by $n$ and letting $n\rightarrow \infty$, and then dividing by $\frac{1}{2}\log P$ and letting $P\rightarrow \infty$, we have
\begin{align}
2d_1+d_2 \leq 2N-K \label{eq:bound1}
\end{align}
By reversing the roles of the transmitters, we have
\begin{align}
d_1+2d_2 \leq 2N-K \label{eq:bound2}
\end{align} 
Combining \eqref{eq:bound1} and \eqref{eq:bound2}, we have the required bound
\begin{align}
d_1+d_2 \leq \frac{2}{3}(2N-K)
\end{align}

This completes the proof of the converse of Theorem \ref{theo}.
 
\section{Achievable Schemes for Fading Channel Gains}\label{sec:achievability} 
We provide separate achievable schemes for each of the following regimes:
\begin{enumerate}
\item $K\leq N/2$
\item $N/2\leq K \leq N$
\item $N\leq K\leq 4N/3$
\item $4N/3 \leq K\leq 3N/2$
\item $3N/2 \leq K \leq 2N$
\end{enumerate} 

Each scheme described in the following sections can be outlined as follows. We neglect the impact of noise at high SNR. Then, to achieve a certain sum s.d.o.f., $d_s$, we achieve the s.d.o.f.~pair $(d_1,d_2)$ with $d_s=d_1+d_2$. We send $n_1$ symbols $\v_1 = \left(v_{11},\ldots,v_{1n_1}\right)$ and $n_2$ symbols $\v_2=\left(v_{21},\ldots,v_{2n_2}\right)$ from the first and second transmitters, respectively, in $n_B$ slots, such that $d_1 = n_1/n_B$ and $d_2 = n_2/n_B$. Finally, we show that the leakage of information symbols at the eavesdropper is $o(\log P)$. We  however want a stronger guarantee of security, namely,
\begin{align}
\frac{1}{n} I(W_1,W_2;\Z^n) \rightarrow 0
\end{align}      
as $n\rightarrow \infty$. To achieve this, we view the $n_B$ slots described in the scheme as a block and treat the equivalent channel from $\v_1$  and $\v_2$ to $\mathbf{Y}$ and $\mathbf{Z}$ as a memoryless multiple access wiretap channel with $\mathbf{Y}$ being the output at the legitimate receiver and $\Z$ being the output at the eavesdropper. The following sum secure rate is achievable \cite{bagherikaram_motahari_khandani_mac}: 
\begin{align}
\sup (R_1+R_2) \geq I(\mathbf{V};\Y) - I(\mathbf{V};\mathbf{Z}) \label{eq:sum_rate}
\end{align}
where $\V \stackrel{\Delta}{=}  \left\lbrace \v_1, \v_2\right\rbrace $. Using the proposed scheme, $\v_1$ and $\v_2$ can be reconstructed from $\mathbf{Y}$ to within noise distortion. Thus, 
\begin{align}
I(\V;\mathbf{Y}) =& (n_1+n_2) \frac{1}{2}\log P + o(\log P)\label{eq:indep1}
\end{align}
Also, for each scheme, by design
\begin{align}
I(\V;\mathbf{Z}) =& o(\log P)\label{eq:leak1}
\end{align}
Thus, from \eqref{eq:sum_rate}, the achievable sum secure rate in each block is $(n_1+n_2)\frac{1}{2}\log P + o(\log P)$. Since our block contains $n_B$ channel uses, the effective sum  secure rate is
\begin{align}
\sup(R_1+R_2) \geq \left(\frac{n_1+n_2}{n_B}\right) \frac{1}{2}\log P + o(\log P)
\end{align}
Thus, the achievable sum s.d.o.f.~is $\frac{n_1+n_2}{n_B}$, with the stringent security requirement as well.

In the following subsections, we present the achievable scheme for each regime.

\subsection{$K\leq N/2$}
In this regime, the optimal sum s.d.o.f.~is $N$. In our scheme, transmitter $1$ sends $(N-K)$ independent Gaussian symbols $\mathbf{v}_1 \in \mathbb{R}^{N-K}$ while transmitter $2$ sends $K$ independent Gaussian symbols $\mathbf{v}_2 \in \mathbb{R}^{K}$, in one time slot. This can be done by beamforming the information streams at both transmitters to directions that are orthogonal to the eavesdropper's channel. To this end, the transmitted signals are:
\begin{align}
\X_1 = \P_1\mathbf{v}_1\\
\X_2 = \P_2\mathbf{v}_2 
\end{align}
where $\P_1 \in \mathbb{R}^{N\times (N-K)}$ is a matrix whose $(N-K)$ columns span the $(N-K)$ dimensional nullspace of $\G_1$, and $\P_2 \in \mathbb{R}^{N\times K}$ is a matrix with $K$ linearly independent vectors drawn from the $(N-K)$ dimensional nullspace of $\G_2$. This can be done since $K\leq N-K$. The channel outputs are:
\begin{align}
\Y =& [\H_1\P_1\quad \H_2\P_2]\left[\begin{array}{c} 
\mathbf{v}_1 \\
\mathbf{v}_2 
\end{array}\right]+ \N_1\\
\Z =& \N_2
\end{align}
Note that $[\H_1\P_1\quad \H_2\P_2]$ is an $N\times N$ matrix with full rank almost surely, and thus, both $\mathbf{v}_1$ and $\mathbf{v}_2$ can be decoded at the legitimate receiver to within noise variance. On the other hand, they do not appear in the eavesdropper's observation and thus their security is guaranteed.

\subsection{$N/2\leq K \leq N$}  
The optimal sum s.d.o.f.~in this regime is $\frac{2}{3}(2N-K)$. Thus, transmitter $i$ sends $(2N-K)$ Gaussian symbols $\left\lbrace\mathbf{v}_i \in  \mathbb{R}^{2K-N}, \tilde{\v}_i(t)  \in \mathbb{R}^{N-K}, t=1,2,3\right\rbrace $, each drawn independently from $\mathcal{N}(0,\bar{P})$, in $3$ time slots for $i=1,2$, where $\bar{P} = \alpha P$ and $\alpha$ is chosen to satisfy the power constraint. Intuitively, transmitter $i$ sends the $(N-K)$ symbols  $\tilde{\v}_i(t)$ by beamforming orthogonal to the eavesdropper in each time slot $t=1,2,3$. The remaining $(2K-N)$ symbols are sent over $3$ time slots using a scheme similar to the SISO scheme of \cite{jianwei_ulukus_one_hop,pritam_one_hop}. Thus, the transmitted signals at time $t$ are:
\begin{align}
\X_1(t) =& \G_1(t)^\perp \tilde{\v}_1(t) + \P_1(t)\v_1 + \H_1(t)^{-1}\Q(t)\u_1\\
\X_2(t) =& \G_2(t)^\perp \tilde{\v}_2(t) + \P_2(t)\v_2 + \H_2(t)^{-1}\Q(t)\u_2  
\end{align}
where $\G_i(t)^{\perp}$ is an $N\times (N-K)$ full rank matrix with $\G_i(t)\G_i(t)^{\perp} = \mathbf{0}_{N\times(N-K)}$, $\u_i$ is a $(2K-N)$ dimensional vector whose entries are drawn in an i.i.d.~fashion from $\mathcal{N}(0,\bar{P})$, and $\P_i$ and $\Q$ are $N\times (2K-N)$ precoding matrices that will be fixed later. The channel outputs are:
\begin{align}
\Y(t) =& \H_1(t)\G_1(t)^\perp \tilde{\v}_1(t)+\H_1(t)\P_1(t)\v_1  + \H_2(t)\P_2(t)\v_2\nonumber\\&+ \H_2(t)\G_2(t)^\perp \tilde{\v}_2(t)
+ \Q(t)(\u_1+\u_2)+\N_1(t)\\
\Z(t) =& \G_1(t)\P_1(t)\v_1 + \G_2(t)\H_2(t)^{-1}\Q(t)\u_2 \nonumber\\
&+ \G_2(t)\P_2(t)\v_2 + \G_1(t)\H_1(t)^{-1}\Q(t)\u_1 + \N_2(t)
\end{align}
We now choose $\Q(t)$ to be any $N\times (2K-N)$ matrix with full column rank, and choose 
\begin{align}
\P_i(t) = \G_i(t)^T(\G_i(t)\G_i(t)^T)^{-1}(\G_j(t)\H_j(t)^{-1})\Q(t)
\end{align}
where $i,j\in \left\lbrace 1,2\right\rbrace, i\neq j$. It can be verified that this selection aligns $\v_i$ with $\u_j$, $i\neq j$, at the eavesdropper, and this guarantees that the information leakage is $o(\log P)$. On the other hand, the legitimate receiver decodes the desired signals $\left\lbrace \tilde{\v}_i(t) \in \mathbb{R}^{N-K},  t\in\left\lbrace 1,2,3\right\rbrace\right\rbrace$, $\left\lbrace \v_i \in \mathbb{R}^{2K-N}, i=1,2 \right\rbrace $ and the aligned artificial noise symbols $\u_1+\u_2 \in  \mathbb{R}^{2K-N} $, i.e., $6(N-K)+3(2N-K)=3N$ symbols using $3N$ observations in $3$ time slots, to within noise variance. This completes the scheme for the regime $N/2\leq K \leq N$.

\subsection{$N\leq K \leq 4N/3$}
In this regime, the optimal sum s.d.o.f.~is $\frac{2}{3}N$. Therefore, transmitter $i$ in our scheme sends $N$ Gaussian symbols, $\v_{i} \in \mathbb{R}^{N}$, in $3$ time slots. The transmitted signals in time slot $t$ are given by
\begin{align}
\X_1(t) = \P_1(t)\v_1 + \H_1(t)^{-1}\Q(t)\u_1\\
\X_2(t) = \P_2(t)\v_2 + \H_1(t)^{-1}\Q(t)\u_2 
\end{align} 
where the $\P_1(t)$, $\Q(t)$, and $ \P_2(t)$ are $N\times N$ precoding matrices to be designed. Let us define
\begin{align}
\tilde{\P}_i \stackrel{\Delta}{=} \left[\begin{array}{c}
\P_i(1)\\
\P_i(2)\\
\P_i(3)
\end{array}\right], \quad 
\tilde{\Q}\stackrel{\Delta}{=}  \left[\begin{array}{c}
\Q(1)\\
\Q(2)\\
\Q(3)
\end{array}\right] 
\end{align}
Further, if we define
\begin{align}
\tilde{\H}_i  \stackrel{\Delta}{=}  \left[\begin{array}{ccc}
\H_i(1) & \mathbf{0}_{N\times N} & \mathbf{0}_{N\times N}\\
\mathbf{0}_{N\times N}& \H_i(2) & \mathbf{0}_{N\times N}\\
\mathbf{0}_{N\times N}&\mathbf{0}_{N\times N}&\H_i(3)
\end{array}\right]
\end{align}
and $\tilde{\G}_i$ similarly, we can compactly represent the channel outputs over all $3$ time slots as
\begin{align}
\tilde{\Y} =& \tilde{\H}_1\tilde{\P}_1\v_1 + \tilde{\H}_2\tilde{\P}_2\v_2 + \tilde{\Q}(\u_1+\u_2) +\tilde{\N}_1\\
\tilde{\Z} =& \tilde{\G}_1\tilde{\P}_1\v_1 + \tilde{\G}_2\tilde{\H}_2^{-1}\tilde{\Q}\u_2 + \tilde{\G}_2\tilde{\P}_2\v_2 + \tilde{\G}_1\tilde{\H}_1^{-1}\tilde{\Q}\u_1 + \tilde{\N}_2  
\end{align}  
where $\tilde{\N}_i \stackrel{\Delta}{=} [\N_i(1)^T\quad \N_i(2)^T\quad\N_i(3)^T ]^T$, $\tilde{\Y}  \stackrel{\Delta}{=} [\Y(1)^T\quad \Y(2)^T\quad\Y(3)^T ]^T$, and $\tilde{\Z}$ is defined similarly. To ensure secrecy, we impose the following conditions
\begin{align}
\tilde{\G}_1\tilde{\P}_1 =& \tilde{\G}_2\tilde{\H}_2^{-1}\tilde{\Q}\label{eq:alignment11}\\
\tilde{\G}_2\tilde{\P}_2=& \tilde{\G}_1\tilde{\H}_1^{-1}\tilde{\Q}\label{eq:alignment2}
\end{align}
We rewrite the conditions in \eqref{eq:alignment11}-\eqref{eq:alignment2} as
\begin{align}
\mathbf{\Psi}\left[\begin{array}{c}
\tilde{\P}_1\\
\tilde{\P}_2\\
\tilde{\Q}
\end{array}\right] = \mathbf{0}_{6K\times N}
\end{align}
where 
\begin{align}
\mathbf{\Psi} \stackrel{\Delta}{=} \left[\begin{array}{ccc}
\tilde{\G}_1 & \mathbf{0}_{3K\times 3N} &  -\tilde{\G}_2\tilde{\H}_2^{-1}\\
\mathbf{0}_{3K\times 3N} & \tilde{\G}_2 & -\tilde{\G}_1\tilde{\H}_1^{-1}
\end{array}\right]
\end{align}
Note that $\mathbf{\Psi}$ has a nullity $9N-6K$. Since $9N-6K\geq N$ in this regime, we can choose $N$ vectors of dimension $9N$ randomly such that they are linearly independent and lie in the nullspace of $\mathbf{\Psi}$. We can then assign to $\tilde{\P}_1$, $\tilde{\P}_2$ and $\tilde{\Q}$, the top, the middle and the bottom $3N$ rows of the matrix comprising the $N$ chosen vectors. This guarantees secrecy of the message symbols at the eavesdropper. 

To see the decodability, we rewrite the received signal at the legitimate receiver as
\begin{align}
\tilde{\Y} =\mathbf{\Phi}\left[\begin{array}{c}
\v_1\\
\v_2\\
\u_1+\u_2
\end{array}\right] + \tilde{\N}_1
\end{align} 
where $\mathbf{\Phi}\stackrel{\Delta}{=}[\tilde{\H}_1\tilde{\P}_1\quad \tilde{\H}_2\tilde{\P}_2\quad \tilde{\Q}]$. We note that $\mathbf{\Phi}$ is $3N\times 3N$ and full rank almost surely; thus, the desired signals $\v_1$ and $\v_2$ can be decoded at the legitimate receiver within noise distortion at high SNR.  

\subsection{$4N/3\leq K \leq 3N/2$}
The optimal s.d.o.f.~in this regime is $2N-K$. To achieve this s.d.o.f., the first transmitter sends $K-N$ Gaussian symbols $\left\lbrace \v_1 \in \mathbb{R}^{3N-2K}, \tilde{\v} \in \mathbb{R}^{3K-4N} \right\rbrace $, while the second transmitter sends $3N-2K$ Gaussian symbols $\left\lbrace \v_2 \in \mathbb{R}^{3N-2K}\right\rbrace $, in one time slot. The scheme is as follows. The transmitted signals are
\begin{align}
\X_1 = \R_1\tilde{\v} + \P_1\v_1 + \H_1^{-1}\Q\u_1\\
\X_2 = \R_2\tilde{\u} + \P_2\v_2 + \H_2^{-1}\Q\u_2
\end{align} 
where $\tilde{\u}\in \mathbb{R}^{3K-4N}$ and $\u_1,\u_2 \in  \mathbb{R}^{3N-2K}$ are artificial noise vectors, whose entries are drawn in an i.i.d.~fashion from $\mathcal{N}(0,\bar{P})$. The precoding matrices $\R_i \in \mathbb{R}^{N\times (3K-4N)}$, and $\P_i,\Q_i \in  \mathbb{R}^{N\times (3N-2K)}$ will be chosen later.  The channel outputs are
\begin{align}
\Y =& \H_1\R_1\tilde{\v} + \H_1\P_1\v_1 + \H_2\P_2\v_2 + \H_2\R_2\tilde{\u} + \Q(\u_1+\u_2)+ \N_1\\
\Z =& \G_1\R_1\tilde{\v} + \G_2\R_2\tilde{\u} + \G_1\P_1\v_1 + \G_2\H_2^{-1}\Q\u_2 + \G_2\P_2\v_2 + \G_1\H_1^{-1}\Q\u_1+\N_2
\end{align}
To ensure secrecy, we want to impose the following conditions:
\begin{align}
\G_1\R_1 =&  \G_2\R_2\label{eq:align1}\\
\G_1\P_1 =&\G_2\H_2^{-1}\Q \label{eq:align21}\\
\G_2\P_2=&\G_1\H_1^{-1}\Q\label{eq:align22}
\end{align}
To satisfy \eqref{eq:align1}, we choose $\R_1$ and $\R_2$ to be the first and the last $N$ rows of a $2N\times 3K-4N$ matrix whose columns consist of any $3K-4N$ linearly independent vectors drawn randomly from the nullspace of $[\G_1\quad -\G_2]$. This is possible since, $3K-4N\leq 2N-K$ in this regime. To satisfy \eqref{eq:align21}-\eqref{eq:align22}, we let $\P_1$, $\P_2$ and $\Q$ to be the first, the second and the last $N$ rows of a $3N\times (3N-2K)$ matrix whose columns are randomly chosen to span the $(3N-2K)$ dimensional nullspace of the matrix $\mathbf{\Lambda}$ given by  
\begin{align}
\mathbf{\Lambda} \stackrel{\Delta}{=} \left[\begin{array}{ccc}
\G_1 & \mathbf{0}_{K\times N}& -\G_2\H_2^{-1}\\
\mathbf{0}_{K\times N} & \G_2 &-\G_1\H_1^{-1}
\end{array}\right]
\end{align}  

To see the decodablity, we can rewrite the observation at the legitimate receiver as
\begin{align}
\Y = \mathbf{\Phi}\left[\begin{array}{c} 
\tilde{\v}  \\
\v_1\\
\v_2\\
\tilde{\u}\\
\u_1 + \u_2
\end{array}\right] + \N_1
\end{align}
where $\mathbf{\Phi}$ is the $N\times N$ matrix defined as 
\begin{align}
\mathbf{\Phi} = \left[\H_1\R_1 \quad \H_1\P_1 \quad \H_2\P_2 \quad \H_2\R_2 \quad \Q\right]
\end{align}
Since $\mathbf{\Phi}$ is full rank almost surely, the legitimate receiver can decode its desired symbols $\tilde{\v}, \v_1$, and $\v_2$.   
\subsection{$3N/2\leq K \leq 2N$}  
In this regime, it is clear from Fig.~\ref{fig:sdof_plot} that the multiple access wiretap channel has the same optimal sum s.d.o.f.~as the optimal s.d.o.f.~of the wiretap channel with one helper. Thus, an optimal achievable scheme for the wiretap channel with one helper suffices as the scheme for the multiple access wiretap channel as well. Such an optimal scheme, based on real interference alignment, is provided in \cite{nafea_yener} for the wiretap channel with one helper with fixed channel gains. Here, we provide a scheme based on vector space alignment. 

In order to achieve the optimal sum s.d.o.f.~of $2N-K$ in this regime, the first transmitter sends $2N-K$ independent Gaussian symbols $\v \in \mathbb{R}^{2N-K}$ securely, in one time slot. The second transmitter just transmits artificial noise symbols $\u \in \mathbb{R}^{2N-K} $, whose entries are drawn in an i.i.d.~fashion from $\mathcal{N}(0,\bar{P})$. The transmitted signals are
\begin{align}
\X_1 = \P\v\\
\X_2 = \Q\u
\end{align}
where $\P$ and $\Q$ are $N\times (2N-K)$ precoding matrices to be fixed later. The received signals are
\begin{align}
\Y =& \H_1\P\v + \H_2\Q\u+\N_1\\
\Z =& \G_1\P\v + \G_2\Q\u+\N_2
\end{align}
To ensure security, we wish to ensure that
\begin{align}
\G_1\P = \G_2\Q
\end{align} 
This can be done by choosing $\P$ and $\Q$ to be the top and the bottom $N$ rows of a $2N\times (2N-K)$ matrix whose linearly independent columns are drawn randomly from the nullspace of $[\G_1\quad -\G_2]$. The decodability is ensured by noting that the matrix $[\H_1\P\quad \H_2\Q]$ is full column rank and $2(2N-K)\leq N$ in this regime. 

\section{Achievable Schemes for Fixed Channel Gains}\label{sec:achievability_real} 
We note that the achievable schemes proposed for the fading channel gains in the regimes $K\leq \frac{N}{2}$ and $\frac{4N}{2} \leq K \leq 2N$ are single time-slot schemes and suffice for the fixed channel gains case. However, in the regime $\frac{N}{2}\leq K \leq \frac{4N}{3}$, the schemes for the fading channel gains exploit the diversity of channel gains over three time slots; thus, these schemes cannot be used in the fixed channel gains case. Therefore, we now propose new achievable schemes for this regime. In this regime, the optimal sum s.d.o.f.~is of the form $2\left(d+\frac{l}{3}\right),\, l=0,1,2$, where $d$ is an integer. When $l=0$, the sum s.d.o.f.~is an integer and carefully precoded Gaussian signaling suffices. However, when $l\neq 0$, the s.d.o.f.~has a fractional part, and Gaussian signaling alone is not optimal, since Gaussian signals with full power cannot carry fractional d.o.f.~of information. 

The general structure of our schemes is as follows: We decompose the channel input at each transmitter into two parts: a Gaussian signaling part carrying $d$ (the integer part) d.o.f.~of information securely, and a structured signaling part carrying  $\frac{l}{3}$ (the fractional part) d.o.f.~of information securely. The structure of the Gaussian signals carrying the integer s.d.o.f.~$d$ are the same as that of the corresponding schemes for the fading channel gains. This ensures security at the eavesdropper as well as decodability at the legitimate receiver as long as the structured signals carrying the fractional s.d.o.f.~$\frac{2l}{3}$ from both transmitters can be decoded at the legitimate receiver. The design of the structured signals is motivated from the SISO scheme of \cite{jianwei_ulukus_one_hop}. In fact, when $l=1$, we use the signal structure of the scheme in \cite{jianwei_ulukus_one_hop}, where real interference alignment is used to transmit $\frac{2}{3}$ sum s.d.o.f.~on the SISO multiple access wiretap channel. However, when $l=2$,  a new scheme is required to achieve $\frac{4}{3}$ sum s.d.o.f.~on the MIMO multiple access wiretap channel with two antennas at every terminal. To that end, we first provide a novel scheme, based on asymptotic real interference alignment \cite{real_inter_align,real_inter_align_exploit}, for the canonical $2\times2\times2\times2$ MIMO multiple access wiretap channel. 

\subsection{Scheme for the $2\times2\times2\times2$ System}  
The optimal sum s.d.o.f.~is $\frac{4}{3}$. Since the legitimate receiver has 2 antennas, we achieve $\frac{2}{3}$ s.d.o.f.~on each antenna. The scheme is as follows.

Let $m$ be a large integer. Define $M\stackrel{\Delta}{=}m^\Gamma$, where $\Gamma$ will be specified later. The channel inputs are given by
\begin{align}
\X_1=& \G_1^{-1}\G_2\H_2^{-1}\left(\begin{array}{c}\mathbf{t}_1^T\mathbf{v}_{11}\\\mathbf{t}_2\mathbf{v}_{12}\end{array}\right) + \H_1^{-1}\left(\begin{array}{c}\mathbf{t}_1^T\mathbf{u}_{11}\\\mathbf{t}_2\mathbf{u}_{12}\end{array}\right)\\
\X_2=& \G_2^{-1}\G_1\H_1^{-1}\left(\begin{array}{c}\mathbf{t}_1^T\mathbf{v}_{21}\\\mathbf{t}_2\mathbf{v}_{22}\end{array}\right) + \H_2^{-1}\left(\begin{array}{c}\mathbf{t}_1^T\mathbf{u}_{21}\\\mathbf{t}_2\mathbf{u}_{22}\end{array}\right)
\end{align}
where $\mathbf{t}_i, i=1,2$ are $M$ dimensional precoding vectors which will be fixed later, and $\mathbf{u}_{ij}, {\v}_{ij}$ are independent random variables drawn uniformly from the same PAM constellation $C(a,Q)$ given by
\begin{align}
C(a,Q) = a\left\lbrace -Q, -Q+1, \ldots, Q-1,Q\right\rbrace 
\end{align} 
where $Q$ is a positive integer and $a$ is a real number used to normalize the transmission power. The exact values of $a$ and $Q$ will be specified later. The variables $\v_{ij}$ denote the information symbols of transmitter $i$, while $\u_{ij}$ are the cooperative jamming signals being transmitted from transmitter $i$.  

The channel outputs are given by
\begin{align}
\Y =& \mathbf{A}\left(\begin{array}{c}\mathbf{t}_1^T\mathbf{v}_{11}\\\mathbf{t}_2\mathbf{v}_{12}\end{array}\right) + \mathbf{B}\left(\begin{array}{c}\mathbf{t}_1^T\mathbf{v}_{21}\\\mathbf{t}_2\mathbf{v}_{22}\end{array}\right) +  \left(\begin{array}{c}\mathbf{t}_1^T(\mathbf{u}_{11}+\u_{21})\\\mathbf{t}_2(\mathbf{u}_{12}+\u_{22})\end{array}\right) + \N_1\label{eq:received_2times2}\\
\Z =& \G_1\H_1^{-1} \left(\begin{array}{c}\mathbf{t}_1^T(\mathbf{u}_{11}+\v_{21})\\\mathbf{t}_2(\mathbf{u}_{12}+\v_{22})\end{array}\right) + \G_2\H_2^{-1}\left(\begin{array}{c}\mathbf{t}_1^T(\mathbf{u}_{21}+\v_{11})\\\mathbf{t}_2(\mathbf{u}_{22}+\v_{12})\end{array}\right)+\N_2
\end{align} 
where $\mathbf{A} = \H_1\G_1^{-1}\G_2\H_2^{-1}$ and $\mathbf{B} = \H_2\G_2^{-1}\G_1\H_1^{-1}$. Note that the information symbols $\v_{ij}$ are buried in the cooperative jamming signals $\u_{kj}$, where $k\neq i$, at the eavesdropper. Intuitively, this ensures security of the information symbols at the eavesdropper. At the legitimate receiver, we can express the received signal $\Y$ more explicitly as
\begin{align}
\left(\begin{array}{c}
\mathbf{t}_2^T(a_{12}\v_{12} + b_{12}\v_{22}) + \mathbf{t}_1^T(a_{11}\v_{11} + b_{11}\v_{21} + \u_{11} + \u_{21})\\
\mathbf{t}_1^T(a_{21}\v_{11} + b_{21}\v_{21}) + \mathbf{t}_2^T(a_{22}\v_{12} + b_{22}\v_{22} + \u_{12}+ \u_{22})
\end{array}\right)
\end{align} 

We define
\begin{align}
T_1 = \left\lbrace a_{11}^{r_1}b_{11}^{r_2}, r_i \in \left\lbrace 0,\ldots,m-1\right\rbrace\right\rbrace  \\
T_2 = \left\lbrace a_{22}^{r_1}b_{22}^{r_2}, r_i \in \left\lbrace 0,\ldots,m-1\right\rbrace  \right\rbrace 
\end{align}
Letting $\Gamma =2$, we note that 
\begin{align}
|T_1| = |T_2| = M
\end{align}   
We choose $\mathbf{t}_i$ to be the $M$ dimensional vector that has all the elements of $T_i$. We note that all elements in $T_i$ are rationally independent, since the channel gains are drawn independently from a continuous distribution. Also, the elements of $T_i$ can be verified to be rationally independent of the elements of $T_j$, if $i\neq j$. With the above selections, let us analyze the structure of the received signal at the legitimate receiver.

At the first antenna, $\u_{11}$ and $\u_{21}$ arrive along the dimensions of $T_1$. The signals $\v_{11}$ and $\v_{21}$ arrive along dimensions $a_{11}T_1$ and $b_{11}T_1$ and, thus, they align with $\u_{11}$ and $\u_{21}$ in $\tilde{T}_1$, where,
\begin{align}
\tilde{T}_1 = \left\lbrace a_{11}^{r_1}b_{11}^{r_2}, r_i \in \left\lbrace 0,\ldots,m\right\rbrace\right\rbrace 
\end{align}
Thus, $\v_{11}$ and $\v_{21}$ cannot be reliably decoded from the observation of the first antenna. However, the desired signals $\v_{12}$ and $\v_{22}$ arrive along dimensions $a_{12}T_2$ and $b_{12}T_2$, respectively. Note that the elements of $a_{12}T_2$ and $b_{12}T_2$ are rationally independent and thus, $\v_{12}$ and $\v_{22}$ occupy separate rational dimensions. Also they are separate from the interference space $\tilde{T}_1$. Therefore, $\v_{12}$ and $\v_{22}$ can be reliably decoded at high SNR. Heuristically, the s.d.o.f.~achieved using the first antenna is $\frac{2|T_1|}{2|T_1|+|\tilde{T}_2|} = \frac{2m^2}{2m^2+(m+1)^2} \approx \frac{2}{3}$ for large enough $m$.  

At the second antenna, a similar analysis holds. The signals $\v_{12}$, $\v_{22}$, $\u_{12}$ and $\u_{22}$ align with each other in the dimensions of $\tilde{T}_2$, which is defined as   
\begin{align}
\tilde{T}_2 = \left\lbrace a_{22}^{r_1}b_{22}^{r_2}, r_i \in \left\lbrace 0,\ldots,m\right\rbrace  \right\rbrace 
\end{align}
The signals $\v_{11}$ and $\v_{21}$ arrive along dimensions that are separate from each other as well as from the dimensions in $\tilde{T}_2$, and thus, can be decoded reliably. The s.d.o.f.~achieved in the second antenna is also $ \frac{2m^2}{2m^2+(m+1)^2} \approx  \frac{2}{3}$ for large $m$. Therefore, the sum s.d.o.f.~achieved using both antennas is $\frac{4}{3}$, as desired. 

Formally, an achievable sum rate is given in  equation \eqref{eq:sum_rate}, where $\mathbf{V} \stackrel{\Delta}{=} \left\lbrace \v_{ij}, i,j \in \left\lbrace 1,2 \right\rbrace \right\rbrace $. 
In order to bound the term $I(\V;\Y)$, we first bound the probability of error. Let $M_S \stackrel{\Delta}{=} 2m^2+(m+1)^2$ be the number of rational dimensions at each receiver antenna. Also let $\V_i = \left\lbrace \v_{kj}, k=1,2; j\neq i \right\rbrace $ be the desired symbols at the $i$th antenna of the receiver. In order to decode, the receiver makes an estimate $\hat{V}_i$ of $\V_i$ by choosing the closest point in the constellation based on the signal received at antenna $i$. For any $\delta>0$, there exists a positive constant
$\gamma$, which is independent of $P$, such that if we choose $Q =
P^{\frac{1-\delta}{2(M_S+\delta)}}$ and $a = \frac{\gamma
	P^{\frac{1}{2}}}{Q}$,
then for almost all channel gains the average power constraint is satisfied
and the probability of error, $\mathrm{Pr}(\V_i\neq \hat \V_i)$, is upper-bounded by $\exp\left( - \eta_{\gamma} P^{{ \delta}} \right)$,
where $\eta_{\gamma}$ is a positive constant which is
independent of $P$. Since $\V = \left\lbrace \V_i, i=1,2\right\rbrace$,
\begin{align}
\mathrm{Pr}(\V\neq \hat \V) \le 2\exp\left( - \eta_{\gamma} P^{{ \delta}} \right)
\end{align} 
By Fano's inequality and the Markov chain $\V\rightarrow \Y \rightarrow \hat{\V}$,
\begin{align}
I(\V; \Y) & = H(\V) - H(\V|\hat{\V}) \\
& \ge \log( |\mathcal{V}| )- 1 - \mathrm{Pr}(\V\neq \hat \V)  \log ( |\mathcal{V}| ) \\
& = \log( |\mathcal{V}| )- o(\log P) \\
& =
\frac{ 4M  (1-\delta)} { M_S + \delta }
\left(\frac{1}{2} \log P\right) + o(\log P) \label{eq:decode}
\end{align}
where $\mathcal{V}$ is the alphabet of $\V$ with cardinality $(2Q+1)^{4M} = (2Q+1)^{4m^2}$. Next, we compute 
\begin{align}
I(\mathbf{V};\mathbf{Z}) \leq&  I\left(\left\lbrace\v_{ij}, i,j=1,2\right\rbrace; \left\lbrace \v_{ij}+\u_{\hat{i}j}, \begin{array}{c}
\hat{i}\neq i,\\
i,j=1,2 
\end{array} \right\rbrace  \right)\\
\leq& \sum_{i,j=1,\hat{i}\neq i}^2 H(\v_{ij}+\u_{\hat{i}j}) - H(\u_{\hat{i}j})\\
\leq& 4M\log(4Q+1) - 4M\log(2Q+1)\\
\leq& 4M  = o(\log P) \label{eq:secrecy}
\end{align}

Using \eqref{eq:decode} and \eqref{eq:secrecy} in \eqref{eq:sum_rate}, we have
\begin{align}
\sup (R_1+R_2) \geq& \frac{ 4M  (1-\delta)} { M_S + \delta }
\left(\frac{1}{2} \log P\right) + o(\log P)
\end{align}
By choosing $\delta$ small enough and $m$ large enough, we can make the sum s.d.o.f.~arbitrarily close to $\frac{4}{3}$.
\subsection{Achievable Schemes for $\frac{N}{2}\leq K \leq N$}

We use structured PAM signaling along with Gaussian signaling. Let $d= \lfloor \frac{2K-N}{3} \rfloor$, and $l=(2K-N)\mbox{mod } 3= (2N-K)\mbox{mod }3$.  Let $\v_i^{(1)} = \left\lbrace v_{ij}, j=1,\ldots,d \right\rbrace$, where each $v_{ij}, j=1,\ldots,d$ is drawn in an i.i.d.~fashion $\sim \mathcal{N}(0,\alpha P) $, and $\v_i^{(2)}=\left\lbrace v_{i(d+1)},\ldots,v_{i(d+l)}\right\rbrace $ are structured PAM signals to be specified later. When $l=0$, $\v_i^{(2)}$ is the empty set. Let $\v_i= \left( \v_i^{(1)}, \v_i^{(2)}\right)$. Also, let $\tilde{\v}_i = \left\lbrace \tilde{v}_{ij}, j=1,\ldots,N-K \right\rbrace$ denote the symbols that can be transmitted securely by beamforming orthogonal to the eavesdropper channel. Transmitter $i$ sends:
\begin{align}
\X_i =& \G_i^\perp \tilde{\v}_i + \P_i\v_i + \H_i^{-1}\Q\u_i 
\end{align}
where $\G_i^{\perp}$ is an $N\times (N-K)$ full rank matrix with $\G_i\G_i^{\perp} = \mathbf{0}_{N\times(N-K)}$, $\u_i = \left( \u_i^{(1)}, \u_i^{(2)}\right) $ is a $(d+l)$ dimensional vector with the entries of $\u_i^{(1)} = \left\lbrace u_{ij}, j=1,\ldots,d\right\rbrace $ being drawn independently of $\v$ and each other from $\mathcal{N}(0,\alpha P)$, and the structure of $\u_i^{(2)}= \left\lbrace u_{i(d+1)},\ldots,u_{i(d+l)}\right\rbrace $ will be specified later. $\P_i$ and $\Q$ are $N\times (d+l)$ precoding matrices that will also be fixed later. The received signals are:
\begin{align}
\Y =& \H_1\G_1^\perp \tilde{\v}_1+\H_1\P_1\v_1  + \H_2\P_2\v_2+ \H_2\G_2^\perp \tilde{\v}_2
+ \Q(\u_1+\u_2)+\N_1\\
\Z =& \G_1\P_1\v_1 + \G_2\H_2^{-1}\Q\u_2 + \G_2\P_2\v_2 + \G_1\H_1^{-1}\Q\u_1+\N_2
\end{align}
We now choose $\Q$ to be any $N\times (d+l)$ matrix with full column rank, and choose $\P_i = \G_i^T(\G_i\G_i^T)^{-1}(\G_j\H_j^{-1})\Q$,
where $i,j\in \left\lbrace 1,2\right\rbrace, i\neq j$. It can be verified that this selection aligns $\v_i$ with $\u_j$, $i\neq j$, at the eavesdropper, and this guarantees that the information leakage is $o(\log P)$. Next, let $\P_i^{(1)}$, $\Q^{(1)}$ be matrices containing the first $d$ columns of $\P_i$ and $\Q$, respectively, while $\P_i^{(2)}$ and $\Q^{(2)}$ contain the last $l$ columns of $\P_i$ and $\Q$, respectively. Let $\mathbf{B}$ be a matrix whose columns lie in the nullspace of the matrix $\mathbf{F}^T = [\H_1\G_1^\perp\quad\H_2\G_2^\perp\quad \H_1{\P}_1^{(1)} \quad \H_1{\P}_1^{(1)}\quad {\Q}^{(1)}]^T$. Note that $\mathbf{F}$ is a $(N-l)\times N$ matrix and thus there exists a $N\times l$ matrix $\mathbf{B}$ such that $\mathbf{FB}=\mathbf{0}$. We consider the filtered output $[\tilde{\Y}, \hat{\Y}]^T = \mathbf{E}\Y$, where
\begin{align}
\mathbf{E} = \left(\begin{array}{cc}
\multicolumn{2}{c}{\mathbf{D}_{l\times N}}\\
\mathbf{I}_{N-l} & \mathbf{0}_{(N-l)\times l}
\end{array}\right)\label{eq:filter}
\end{align}
and $\mathbf{D}=(\mathbf{B}^T\Q^{(2)})^{-1}\mathbf{B}^T$ and let
\begin{align}
\tilde{\Y} =& \mathbf{D}\H_1\P_{1}^{(2)}\v_{1}^{(2)}  + \mathbf{D}\H_2\P_{2}^{(2)}\v_{2}^{(2)}+ (\u_{1}^{(2)}+\u_{2}^{(2)})+\mathbf{D}\N_1\label{eq:filtered1}
\end{align}  
Note that \eqref{eq:filtered1} represents the output at the receiver of a multiple access wiretap channel with $l$ antennas at each terminal. If $l=1$, we let $\v_{i}^{(2)}=v_{i(d+1)}$ be drawn uniformly and independently from the PAM constellation $C(a,Q)$, with $Q =
P^{\frac{1-\delta}{2(3+\delta)}}$ and $a = \frac{\gamma P^{\frac{1}{2}}}{Q}$. Also, $\u_i^{(2)}=u_{i(d+1)}$ is chosen uniformly from $C(a,Q)$ and independently from $\v_j, j=1,2$. The receiver can then decode $v_{1(d+1)}$, $v_{2(d+1)}$ and $(u_{1(d+1)}+u_{2(d+1)})$ with vanishing probability of error. On the other hand, if $l=2$, we choose $\v_i^{(2)}$ and $\u_i^{(2)}$ as in the $2\times2\times2\times2$ multiple access wiretap channel, i.e., $v_{i(d+k)}=\mathbf{t}_{k}^T\hat{\v}_{ik}, k=1,2$, where $\hat{\v}_{ik}$ is an $M$ dimensional vector whose entries are drawn from the PAM constellation $C(a,Q)$ with $Q =P^{\frac{1-\delta}{2(M_S+\delta)}}$ and $a = \frac{\gamma
	P^{\frac{1}{2}}}{Q}$, and $\mathbf{t}_i$ is chosen appropriately analogous to the selection for the $2\times2\times2\times 2$ multiple access wiretap channel, noting the similarity of \eqref{eq:filtered1} with \eqref{eq:received_2times2}. The cooperative jamming signal $\u_i^{(2)}$ is chosen similarly. Then, the receiver can decode $\v_i^{(2)}$ and also $\u_1^{(2)}+ \u_2^{(2)}$ with vanishing probability of error.

Thus, for $l=1,2$, $\v_i^{(2)}$ and $\u_1^{(2)}+ \u_2^{(2)}$ can be eliminated from $\hat{\Y}$. Noting that $2(N-K) + 3d \leq N-l$, $\tilde{\v}_i$ and $\v_i^{(1)}$ can also be decoded from $\tilde{\Y}$. We compute 
\begin{align}
I(\v_1,\v_2, \tilde{\v}_1,\tilde{\v}_2;\Y) =& I(\v_1^{(1)},\v_2^{(1)},\tilde{\v}_1,\tilde{\v}_2;\Y|\v_{1}^{(2)}, \v_{2}^{(2)}) + I(\v_{1}^{(2)}, \v_{2}^{(2)};\Y)\label{eq:decode1}
\end{align}
The second term depends on the value of $l$. When $l=1$, 
\begin{align}
I(\v_{1}^{(2)}, \v_{2}^{(2)};\Y) =& \log(2Q+1)^{2}  + o(\log P)\\
=& 2\frac{1-\delta}{(3+\delta)} \left( \frac{1}{2}\log P\right)  + o(\log P)\label{eq:decode_siso}
\end{align}
On the other hand, when $l=2$, we have
\begin{align}
I(\v_{1}^{(2)}, \v_{2}^{(2)};\Y) =& \frac{ 4M  (1-\delta)} { M_S + \delta }
\left(\frac{1}{2} \log P\right) + o(\log P) \label{eq:decode_mimo}
\end{align}
Thus, in either case, by choosing $\delta$ sufficiently small and $m$ large enough when $l=2$, we have
\begin{align}
I(\v_{1}^{(2)}, \v_{2}^{(2)};\Y) = \frac{2l}{3} \left(\frac{1}{2} \log P\right) + o(\log P) \label{eq:term1}
\end{align}
Noting that $\v_1^{(1)},\v_2^{(1)},\tilde{\v}_1,\tilde{\v}_2$ can be decoded to within noise variance from $\Y$, given $\v_{1}^{(2)}, \v_{2}^{(2)}$, the first term of \eqref{eq:decode1} is
\begin{align}
I(\v_1^{(1)},\v_2^{(1)},\tilde{\v}_1,\tilde{\v}_2;\Y|\v_{1}^{(2)}, \v_{2}^{(2)})\geq & 2(d+ N-K)\left( \frac{1}{2}\log P\right)  + o(\log P)\label{eq:term2}
\end{align} 
Using \eqref{eq:term1} and \eqref{eq:term2} in \eqref{eq:decode1}, we have,
\begin{align}
I(\v_1,\v_2, \tilde{\v}_1,\tilde{\v}_2;\Y) \geq& 2\left(d+N-K+\frac{l}{3}\right)\left( \frac{1}{2}\log P\right)  + o(\log P)\\
=& \frac{2}{3}\left(2N-K\right)\left( \frac{1}{2}\log P\right)  + o(\log P)
\end{align}
This completes the achievable schemes for the regime $\frac{N}{2}\leq K \leq N$.

\subsection{Achievable Schemes for $N\leq K \leq \frac{4N}{3}$}
As in the previous regime, we use structured PAM signaling along with Gaussian signaling. Let $d= \lfloor \frac{N}{3} \rfloor$ and $l=N \mbox{mod }3$. Let $\v_{i} = \left(\v_i^{(1)}, \v_i^{(2)} \right) $ be the information symbols such that the entries of $\v_{i}^{(1)} = \left\lbrace v_{ij}, j=1,\ldots,d\right\rbrace $ are drawn in an i.i.d.~fashion $\sim \mathcal{N}(0,\alpha P)$, and the entries of $\v_i^{(2)} = \left\lbrace v_{ij}, j=d+1,\ldots,d+l\right\rbrace $ are structured PAM signals to be designed later. Let $\u_i = \left(\u_{i}^{(1)}, \u_i^{(2)}\right)$ denote the cooperative jamming symbols such that the entries of $\u_i^{(1)}= \left\lbrace u_{ij}, j=1,\ldots,d\right\rbrace $ are drawn in an i.i.d.~fashion $\sim \mathcal{N}(0,\alpha P)$, and the entries of $\u_i^{(2)} = \left\lbrace u_{ij}, j=d+1,\ldots,d+l\right\rbrace $ are structured PAM signals independent of $\v_j, j=1,2$ and $\u_j, j\neq i$. Transmitter $i$ sends
\begin{align}
\X_i = \P_i\v_i + \H_i^{-1}\Q\u_i 
\end{align} 
where the $\P_1$, $\Q$, and $ \P_2$ are $N\times (d+l)$ precoding matrices to be designed. 
The channel outputs are given by
\begin{align}
\Y =& \H_1\P_1\v_1 + \H_2\P_2\v_2 + \Q(\u_1+\u_2)+\N_1\\
\Z =& \G_1\P_1\v_1 + \G_2\H_2^{-1}\Q\u_2 + \G_2\P_2\v_2 + \G_1\H_1^{-1}\Q\u_1 +\N_2 
\end{align}  
To ensure secrecy, we impose that for $i\neq j$
\begin{align}
\G_i\P_i =& \G_j\H_j^{-1}\Q\label{eq:alignment1}
\end{align}
We rewrite the conditions in \eqref{eq:alignment1} as
\begin{align}
\mathbf{\Psi}\left[\begin{array}{ccc}
\P_1^T & \P_2^T & \Q^T
\end{array}\right]^T = \mathbf{0}_{2K\times (d+l) }
\end{align}
where 
\begin{align}
\mathbf{\Psi} \stackrel{\Delta}{=} \left[\begin{array}{ccc}
\G_1 & \mathbf{0}_{K\times N} &  -\G_2\H_2^{-1}\\
\mathbf{0}_{K\times N} & \G_2 & -\G_1\H_1^{-1}
\end{array}\right]
\end{align}
Note that $\mathbf{\Psi}$ has a nullity $3N-2K$. This alignment is feasible if $3N-2K\geq d+l$, i.e., if $K \leq 4d+l$. This is satisfied since, in this regime, $K\leq 4d + l+ \frac{1}{3}l$, which implies $K\leq 4d+1$ for integers $N$ and $K$, since $0\leq l\leq 2$. This guarantees security and the information leakage is $o(\log P)$. Next, let $\P = \left( {\P}_i^{(1)}, {\P}_i^{(2)} \right)$ such that $ {\P}_i^{(1)}, $ contains the first $d$ columns of $\P_i$. We define $\Q^{(1)}$ and $\Q^{(2)}$ similarly. Let $\mathbf{B}$ be a matrix whose columns lie in the nullspace of the matrix $\mathbf{F}^T = [\H_1{\P}_1^{(1)} \quad \H_1{\P}_1^{(1)}\quad {\Q}^{(1)}]^T$. Note that $\mathbf{F}$ is a $(N-l)\times N$ matrix and thus there exists a non-zero $N\times l$ matrix $\mathbf{B}$ such that $\mathbf{FB}=\mathbf{0}$. We consider the filtered output $[\tilde{\Y}, \hat{\Y}]^T = \mathbf{E}\Y$, where $\mathbf{E}$ is as in \eqref{eq:filter}. We have
\begin{align}
\tilde{\Y} =& \mathbf{D}\H_1\P_{1}^{(2)}\v_{1}^{(2)}  + \mathbf{D}\H_2\P_{2}^{(2)}\v_{2}^{(2)}+ (\u_{1}^{(2)}+\u_{2}^{(2)})+\mathbf{D}\N_1\label{eq:filtered2}
\end{align}  
When $l=1$, we choose $\v_i^{(2)}=v_{i(d+1)}$ and $\u_i^{(2)}=u_{i(d+1)}$ to be PAM signals drawn independently from $C(a,Q)$ with $Q =
P^{\frac{1-\delta}{2(3+\delta)}}$ and $a = \frac{\gamma
	P^{\frac{1}{2}}}{Q}$. The receiver can then decode $v_{1(d+1)}$, $v_{2(d+1)}$ and $(u_{1(d+1)}+u_{2(d+1)})$ with vanishing probability of error. When $l=2$, we choose $\v_i^{(2)}$ and $\u_i^{(2)}$ analogous to the case of the $2\times2\times2\times 2$ multiple access wiretap channel, i.e., $v_{i(d+k)}=\mathbf{t}_{k}^T\hat{\v}_{ik}, k=1,2$, where $\hat{\v}_{ik}$ is an $M$ dimensional vector whose entries are drawn from the PAM constellation $C(a,Q)$ with $Q =P^{\frac{1-\delta}{2(M_S+\delta)}}$ and $a = \frac{\gamma
	P^{\frac{1}{2}}}{Q}$, and $\mathbf{t}_i$ is chosen appropriately, noting the similarity of \eqref{eq:filtered2} with \eqref{eq:received_2times2}. The cooperative jamming signals $\u_i^{(2)}, i=1,2$ are chosen similarly. Such a selection allows the receiver to decode $\v_i^{(2)}$ and also $\u_1^{(2)}+ \u_2^{(2)}$ with vanishing probability of error. Thus, they can be eliminated from the received observation $\Y$.

Thus, we can eliminate $\v_i^{(2)}$ and $\u_1^{(2)}+\u_2^{(2)}$ from $\hat{\Y}$. Noting that $3d \leq N-l$, $\v_{i}^{(1)} = \left\lbrace v_{ij}, j=1,\ldots,d\right\rbrace $ can also be decoded to within noise variance from $\Y$. As in \eqref{eq:decode_siso}-\eqref{eq:decode_mimo},
\begin{align}
I(\v_{1}^{(2)}, \v_{2}^{(2)};\Y) =\frac{2l}{3} \left(\frac{1}{2} \log P\right) + o(\log P) \label{eq:term21}
\end{align}
Also, as in \eqref{eq:term2}, we have 
\begin{align}
I(\v_1^{(1)},\v_2^{(1)};\Y|\v_{1}^{(2)}, \v_{2}^{(2)})\geq& 2d\left( \frac{1}{2}\log P\right)  + o(\log P)\label{eq:term22}
\end{align}
Using \eqref{eq:term21} and \eqref{eq:term22}, we have
\begin{align}
I(\v_1,\v_2;\Y) \geq& 2\left(d+\frac{l}{3}\right)\left( \frac{1}{2}\log P\right)  + o(\log P)\\
=& \frac{2}{3}N\left( \frac{1}{2}\log P\right) + o(\log P)
\end{align}

\section{Conclusions}
In this paper, we determined the optimal sum s.d.o.f.~of the two-user MIMO multiple access wiretap channel with $N$ antennas at each transmitter, $N$ antennas at the legitimate receiver and $K$ antennas at the eavesdropper. For the case of fading channel gains, we provided vector space alignment based achievable schemes that exploit the channel variation over multiple time slots in general. When the channel gains are fixed, such channel diversity is not available, and we provided single time-slot  schemes that use real interference alignment on structured signaling. We also provided matching converses to establish the optimality of the achievable schemes for both fixed and fading channel gains. Our results highlight the effect of the number of eavesdropper antennas on the s.d.o.f.~of the multiple access wiretap channel. 
\bibliographystyle{unsrt}
\bibliography{references}
\end{document}